\documentclass{article}
\usepackage{ijcai25}

\usepackage{times}
\usepackage{soul}
\usepackage{url}
\usepackage[hidelinks]{hyperref}
\usepackage[utf8]{inputenc}
\usepackage[small]{caption}
\usepackage{graphicx}
\usepackage{amsmath}
\usepackage{amsthm}
\usepackage{svg}
\usepackage{booktabs}
\usepackage{algorithm}
\usepackage{algorithmic}
\usepackage[switch]{lineno}
\usepackage{color, xcolor}

\title{Divide and Conquer: Multimodal Video Deepfake Detection via Cross-Modal Fusion and Localization}

\author{
Qingcao Li$^{1,2}$\thanks{Equal contribution.}
\and
Miao He$^1$\footnotemark[1]
\and
Liang Yi$^1$
\and
Qing Wen$^1$
\and
Yitao Zhang$^1$
\and
Hongshuo Jin$^1$
\and
Peng Cheng$^{1,3}$
\and
Zhongjie Ba$^{1,3}$
\and
Li Lu$^{1,3}$
\And
Kui Ren$^{1,3}$\\
\affiliations
$^1$The State Key Laboratory of Blockchain and Data Security, Zhejiang University\\
$^2$School of Cyber Science and Engineering, Nanjing University of Science and Technology\\
$^3$Hangzhou High-Tech Zone (Binjiang) Institute of Blockchain and Data Security\\
\emails
\{liqingcao@njust.edu.cn, miao\_he@zju.edu.cn, yiliang@zju.edu.cn, qwerwenqing@zju.edu.cn,\newline
12521031@zju.edu.cn, 994195842@qq.com, peng\_cheng@zju.edu.cn, zhongjieba@zju.edu.cn,\newline
li.lu@zju.edu.cn, kuiren@zju.edu.cn\}
}

\begin{document}

\maketitle

\begin{abstract}
This paper presents a system for detecting fake audio-visual content (i.e., video deepfake), developed for Track 2 of the DDL Challenge. The proposed system employs a two-stage framework, comprising unimodal detection and multimodal score fusion. Specifically, it incorporates an audio deepfake detection module and an audio localization module to analyze and pinpoint manipulated segments in the audio stream. In parallel, an image-based deepfake detection and localization module is employed to process the visual modality. To effectively leverage complementary information across different modalities, we further propose a multimodal score fusion strategy that integrates the outputs from both audio and visual modules. Guided by a detailed analysis of the training and evaluation dataset, we explore and evaluate several score calculation and fusion strategies to improve system robustness. Overall, the final fusion-based system achieves an AUC of 0.87, an AP of 0.55, and an AR of 0.23 on the challenge test set, resulting in a final score of 0.5528.

\end{abstract}

\begin{figure}[ht]
\centering
\includegraphics[scale=0.4]{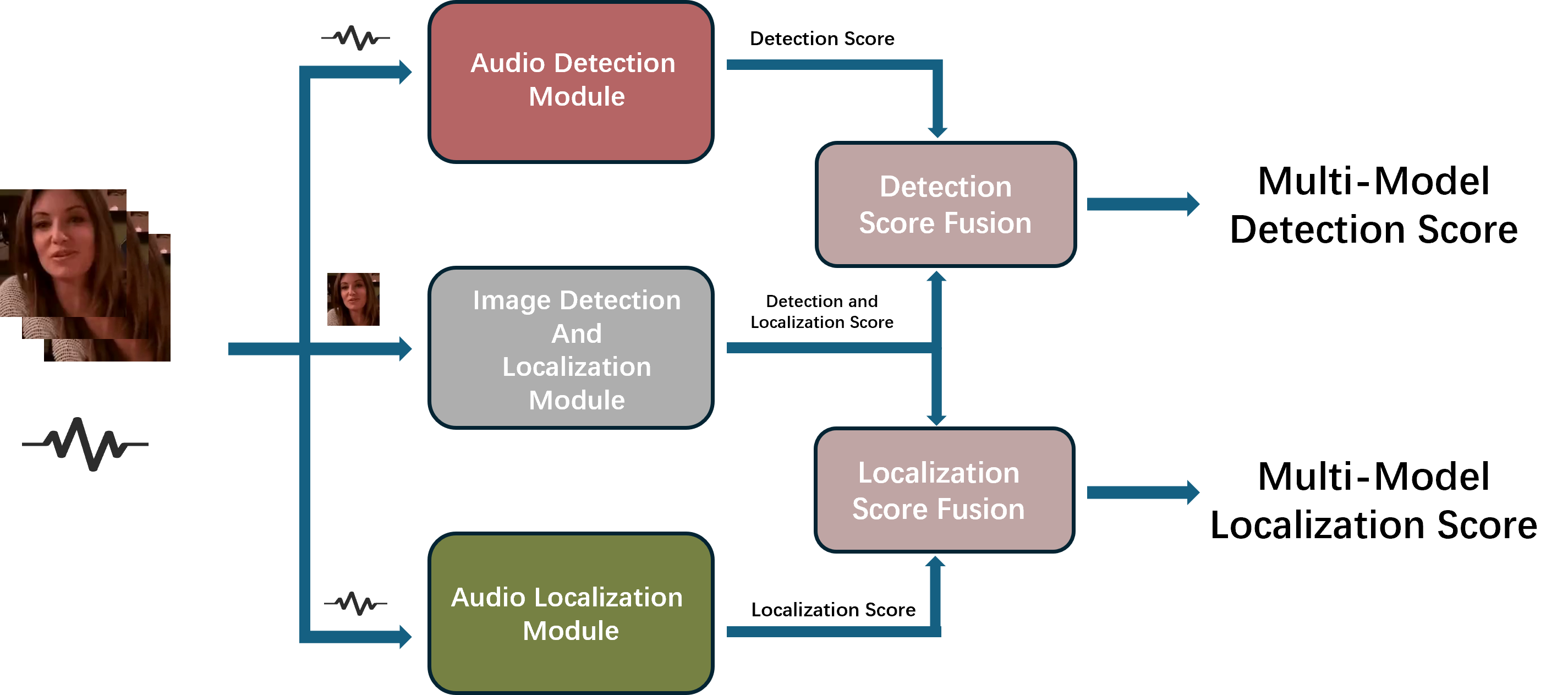}
\caption{
Overview of the proposed audio-visual deepfake detection and localization framework.
The system is composed of three modality-specific modules and two score fusion units.
The \textbf{Audio Detection Module} outputs a single video-level confidence score indicating the presence of audio tampering.
The \textbf{Audio Localization Module} produces frame-level forgery likelihoods, which are aggregated to generate interval-level scores based on pre-defined temporal segments.
The \textbf{Image Detection and Localization Module} concurrently provides both video-level detection and frame-wise visual tampering scores.
For detection, video-level scores from the audio and image branches are fused via the \textbf{Detection Score Fusion}.
For localization, frame- or interval-level outputs from both audio and image streams are aligned and merged by the \textbf{Localization Score Fusion} to generate a unified temporal localization result.
}
\label{fig:overview}
\end{figure}

\section{Introduction}

With the rapid advancement of AI-generated content (AIGC) technologies, including face synthesis/replacement and audio synthesis, the realism and naturalness of synthetic media have reached unprecedented levels~\cite{croitoru2024deepfake}. Although these breakthroughs enable innovative applications in entertainment and human-computer interaction, they also raise serious concerns about the authenticity of information. A growing threat is the proliferation of partially manipulated videos~\cite{liu2024evolving}, where specific frames are selected to synthesize or edit facial expressions or voices and seamlessly embedded in genuine recordings. These multimodal forgeries are particularly difficult to detect, as they often exploit temporal and contextual consistency to evade both human perception and algorithmic detection\cite{kharel2023dftransfusion,tian2023unsupervised,yu2024mining,zhang2023ummaformer}. Therefore,  it is imperative to design robust, generalizable, and multimodal-aware detection and localization systems that can effectively identify and analyze both visual and auditory cues of forgery.

The Deepfake Detection and Localization Challenge (DDL-Challenge)~\cite{miao2025ddl,zhang2024mfms,miao2023multi,miao2024mixture,zhang2024inclusion}  was launched to address these challenges. The challenge consists of two tracks. Track 1 Image Detection and Localization (DDL-I) focuses on image-level forgery detection and spatial localization. Track 2: Audio-Visual Detection and Localization (DDL-AV), aims to promote the development of reliable detection and localization techniques for deepfake content, especially under multi-modal and partially forged scenarios. 

This paper specifically focuses on Track 2. In this track, the organizers provide not only clearly defined tasks but also a comprehensive benchmark dataset~\cite{miao2025ddl} that captures the complexity and subtlety of real-world forgeries. The dataset is divided into training, validation, and test sets, and includes three types of samples: (1) authentic samples, where both audio and video remain untouched; (2) fully forged samples, in which both modalities are synthetically generated throughout the entire video; and (3) Partially forged samples refer to videos where only a subset of the audio and/or visual content is manipulated, rather than the entire sequence. These manipulations may target short segments within either or both modalities, and often involve operations such as insertion, substitution, or blending of synthetic content~\cite{xu2025localization,yan2024generalizing,shiohara2022detecting}. In some cases, forged audio and video fragments may coexist but appear at different times, leading to subtle temporal inconsistencies—either synchronized or intentionally misaligned across modalities. Such localized and temporally dispersed alterations pose significant challenges for detection, as large portions of the original content remain untouched, and the overall coherence of the media may be preserved. The dataset also spans various levels of audio and video quality to reflect real-world recording conditions. Notably, the test set contains a broader range of sophisticated and previously unseen forgery techniques, posing a significant challenge to the generalization and robustness of competing models.

To comprehensively evaluate performance, Track 2 defines two main tasks: detection and localization. The detection task is measured using the Area Under the Receiver Operating Characteristic Curve (AUC)~\cite{hanley1982meaning}, reflecting the model’s ability to distinguish real from fake. Localization performance is evaluated using Average Precision (AP)~\cite{zhang2009average} and Average Recall (AR)~\cite{hosang2015makes}, which assess the accuracy and completeness of tampered segment identification. The final score is computed as the average of detection and localization scores, thereby ensuring a balanced and holistic evaluation framework.

This paper proposes a comprehensive system designed to address the challenge of full and partial multimodal forgeries in videos, focusing on the detection and localization of manipulated audio and visual segments. As illustrated in Figure~\ref{fig:overview}, the system consists of three modality-specific processing modules and two cross-modal fusion components. The audio stream is first processed by two separate modules: the Audio Detection Module~\cite{tak2022automatic}, which outputs a scalar score reflecting the likelihood of audio tampering at the video level, and the Audio Localization Module~\cite{zhong24_interspeech}, which predicts a temporal sequence of forgery probabilities across the audio timeline. Simultaneously, the Image Detection and Localization Module~\cite{chollet2017xception} operates on the visual stream, performing both video-level detection and frame-wise spatial localization of manipulated content. To enhance robustness and leverage the complementary nature of audio and visual cues, we introduce two score-level fusion mechanisms. The Detection Score Fusion aggregates the video-level predictions from the audio and image detection branches, producing a unified deepfake detection score for the entire video. Similarly, the Localization Score Fusion combines the temporal prediction sequence from both audio and image localization module, yielding integrated temporal forgery localization intervals. This hierarchical fusion design allows the system to effectively capture modality-specific inconsistencies while simultaneously modeling their cross-modal interactions, which is crucial for handling partial and asynchronous manipulations.

\section{Data Analysis}\label{sec:data_analysis}

To inform the design of robust forgery detection models, we conducted a systematic analysis of the dataset provided for this competition. Our motivation lies in understanding the characteristics of video data and identifying potential sources of interference that may affect detection performance. The analysis includes two main components:

\begin{itemize}
    \item Statistical analysis of the distribution, proportion, and temporal characteristics of different forgery types;
    \item Manual inspection of a sampled subset, including perceptual assessment of video quality and inspection of forged segments.
\end{itemize}

The dataset comprises a total of 325,975 video samples, with 200860 in the training set, 13,965 in the validation set, and 111,150 in the test set. Table~\ref{tab:category-distribution} shows that within the training and validation sets, authentic videos (real\_audio\_real\_visual) account for approximately 34\% of the total, while fake\_audio\_real\_visual, real\_audio\_fake\_visual, and fake\_audio\_fake\_visual represent roughly 23\%, 24\%, and 19\%, respectively.

\begin{table}[th]
\centering
\resizebox{\linewidth}{!}{
\begin{tabular}{ccc}
\toprule
\textbf{Category} & \textbf{Train + Val Count} & \textbf{Proportion (\%)}\\
\midrule
real\_audio\_real\_visual   & 67,348 + 4,965        & 34\%\\
fake\_audio\_real\_visual   & 46,933 + 3,000        & 23\%\\
real\_audio\_fake\_visual   & 48,035 + 3,000        & 24\%\\
fake\_audio\_fake\_visual   & 38,544 + 3,000        & 19\% \\
\bottomrule
\end{tabular}}
\caption{Distribution of video categories in the training and validation sets.}
\label{tab:category-distribution}
\end{table}

We conducted a statistical analysis of forged segments durations. Across both audio and visual modalities, we categorize forged segments into four temporal intervals. As shown in Table~\ref{tab:segment-duration}, the majority of forged segments are concentrated within the 0-0.5 second range (57\% for audio and 46\% for visual), indicating a typical pattern of short-term perturbations. Furthermore, in the fake\_audio\_fake\_visual category, we observed several cases where audio and visual forgeries overlap.

\begin{table}[th]
\centering
\resizebox{\linewidth}{!}{
\begin{tabular}{ccc}
\toprule
\textbf{Segment Duration (s)} & \textbf{Audio (\% of forged)} & \textbf{Visual (\% of forged)}\\
\midrule
0--0.5  & 57\%  & 46\% \\
0.5--1 & 21\%  & 25\%\\
1--2  & 14\% & 18\%\\
$>$2& 8\% & 11\% \\
\bottomrule
\end{tabular}}
\caption{Distribution of forged segment durations.}
\label{tab:segment-duration}
\end{table}

Additionally, nearly half of the samples contain notable interference factors. These include significant background noise, pure music in the background, overlapping speech from multiple speakers, and minority languages in the audio stream. Visually, some frames exhibit severe blurring or lack discernible facial regions. As both authentic and forged samples are affected by such disturbances, it is essential to enhance model robustness to these unpredictable conditions.

Beyond statistical analysis, we conducted a manual inspection of 200 randomly selected forged videos, which was performed independently by two annotators (i.e., the authors). Approximately 65\% of the forged audio samples exhibited perceptible anomalies, including timbre shifts, abrupt transitions, insertion artifacts, and robotic or mechanical speech cues. Visually, around 90\% of forged samples presented clear anomalies, such as facial distortions, global frame replacements, flickering artifacts, and unnatural color filters. A portion of the samples also displayed audio-visual desynchronization.

In summary, key findings from our data analysis include:

\begin{itemize}
    \item The dataset contains a large proportion of forged videos, with real samples comprising only ~34\% of the train+val set.
    \item Most forged segments—57\% for audio and 46\% for visual—are short in duration ($\leq0.5$s), indicating a localized manipulation pattern.
    \item Distortions and various forms of interference are observed in both authentic and forged samples.
    \item Manual inspection reveals clear perceptual artifacts in a majority of forgeries, but a small portion remains imperceptible.
\end{itemize}

These findings suggest that effective forgery detection models should emphasize short-term anomaly detection and robustness to noisy conditions.
\section{Method}

\subsection{Audio}
\subsubsection{Detection}

We develop an audio-based forgery detection model trained on the official training set using 2-second cropped audio segments with dynamic labeling. To improve robustness, we apply various data augmentation techniques during training. At inference time, a sliding window approach with max-pooling is used to aggregate segment-level scores into a final prediction. Details of the training strategy, data augmentation, and inference process are provided in the following subsections.

\textbf{Training Strategy.} We adopt Wav2Vec2.0-AASIST~\cite{tak2022automatic}, as the backbone of our audio deepfake detection model, which combines the XLS-R~\cite{babu2021xlsrselfsupervisedcrosslingualspeech} (a multilingual Wav2Vec 2.0 variant) as the frontend feature extractor and an AASIST~\cite{9747766} graph neural network as the classifier. During training, we jointly fine-tune XLS-R and train the AASIST module on the official training set. 

During training, audio is cropped into fixed 2-second segments. As shown in Table~\ref{tab:segment-duration}, nearly 92\% of fake segments fall within this duration. The 2s window strikes a balance between temporal resolution and contextual coverage: longer segments may dilute brief spoofed cues (typically 0.5–1s), while shorter ones may lack sufficient context.


During audio pre-processing, we integrate a \textbf{dynamic labeling strategy} with standard waveform cropping to accommodate partial forgery detection. 
Typically, waveforms are randomly cropped to a fixed length, with repetition applied if the original duration is shorter.
If the metadata contains annotated fake segments, the label of each cropped segment is reassigned based on its overlap with forged intervals. Specifically,


\begin{enumerate}
    \item If no forged segments are annotated, the original audio label is retained.
    
    \item If the audio is labeled as fake and its duration is shorter than 2s, it is repeated and cropped to the target length. The label is set to fake (i.e., 1), since all forged content is preserved within the repeated segment.

    \item If the audio is labeled as fake and its duration exceeds 2s, it is randomly cropped. If the cropped segment overlaps any forged interval, it is labeled as fake (i.e., 1); otherwise, it is labeled as real (i.e., 0).
\end{enumerate}

\begin{algorithm}
\caption{Dynamic Labeling of Cropped Audio}
\textbf{Input:} Original waveform $\textbf{w}$; original label $y \in \{0, 1\}$; target length $T$; forged audio segments $\mathcal{M} = \left\{\left[t_s^{(i)}, t_e^{(i)}\right]\right\}_{i=1}^N$, where $0 \leq t_s^{(i)} < t_e^{(i)} \leq T$ \\
\textbf{Parameters:} Crop interval $[t_s, t_e]$, where $0 \leq t_s < t_e \leq T$ \\
\textbf{Output:} Cropped waveform $\textbf{w}'$; updated label $y' \in \{0, 1\}$

\begin{algorithmic}[1]
    \STATE $\textbf{w}', t_s, t_e \gets \texttt{PAD\_AND\_CROP}(\textbf{w}, T)$
    \IF{$\mathcal{M} = \emptyset$}
        \RETURN $\textbf{w}', y$
    \ELSE
        \IF{$\texttt{LENGTH}(\textbf{w}) < T$}
            \RETURN $\textbf{w}', 1$
        \ELSE
            \FORALL{$[t_s^{(i)}, t_e^{(i)}] \in \mathcal{M}$}
                \IF{$t_e > t_s^{(i)}$ \AND $t_s < t_e^{(i)}$}
                    \RETURN $\textbf{w}', 1$
                \ENDIF
            \ENDFOR
            \RETURN $\textbf{w}', 0$
        \ENDIF        
    \ENDIF
\end{algorithmic}
\end{algorithm}



To address the class imbalance introduced by dynamic labeling, the loss function is reweighted based on the bonafide-to-fake ratio observed in the training set. Furthermore, data augmentation is applied prior to waveform cropping and labeling, as detailed in the following section.

\textbf{Data Augmentation.} To enhance model robustness and mitigate the audio quality degradation discussed in Sec~\ref{sec:data_analysis}, we applied multiple data augmentation techniques to the official training data during model training:

\begin{itemize}
    \item Rawboost~\cite{tak2022rawboost}: Consists of two independently applied modules. (1) linear and non-linear convolutive noise to improve robustness against variations from encoding, compression-decompression, and transmission; (2) signal-dependent impulsive noise to simulate acquisition-related distortions such as clipping, device imperfections, synchronization errors, and computational limitations.
    
    \item SpecAugment~\cite{Park_2019}: Applies time and frequency masking to the spectrogram.
    
    \item Noise: 
    A random noise sample from the MUSAN~\cite{snyder2015musan} dataset is added with a SNR uniformly sampled from 13–20 dB.
    
    \item Background music: A random sample from the MUSAN music dataset is added with an SNR uniformly sampled from 13–20 dB.
    
    \item Background speech: A random sample from the MUSAN speech dataset is added with an SNR uniformly sampled from 13–20 dB.
    
\end{itemize}

During training, 50\% of the samples are augmented using one randomly selected method from the aforementioned techniques, each chosen with equal probability. The remaining 50\% remains unaltered.

\textbf{Inference strategy.} During inference, we employ a sliding window inference strategy to evaluate segment-wise authenticity across each audio clip. 
A sliding window of 2 seconds with a 1-second stride is applied, such that each segment overlaps the previous one by 1 second.
If the final segment is shorter than 2 seconds, it is padded with preceding audio to maintain consistent input length. 
The clip-level score is computed via max-pooling over all segment scores, which is effective for detecting partially forged audio by emphasizing the most suspicious segment. 



\subsubsection{Localization}

We adopt a boundary-aware audio forgery localization system, named Boundary-aware Attention Mechanism (BAM)~\cite{zhong24_interspeech}, which integrates frame-level detection and boundary prediction in a unified framework. The system processes variable-length audio inputs during inference while maintaining temporal precision through dedicated boundary modeling. Technical implementation details are outlined below.

\textbf{Training Strategy.} We implement BAM using the WavLM-Large~\cite{WavLM22} self-supervised pre-trained model as the front-end feature extractor, combined with two core modules: (1) Boundary Enhancement Module (BEM) and (2) Boundary Frame-wise Attention Module (BFAM) to learn discriminative features for distinguishing between real and fake frames.

\begin{itemize}
    \item Boundary Enhancement Module: This module extracts both intra-frame and inter-frame information to derive discriminative boundary features. It employs a frame-wise attention block to capture inter-frame correlations and a 1D-ResNet for intra-frame feature learning. The BE module outputs boundary prediction probabilities through a fully-connected layer with a sigmoid function, followed by threshold-based binarization. The threshold was set at 0.5 by the original author.
    
    \item Boundary Frame-wise Attention Module: Leveraging the boundary predictions from the BE module, this component explicitly controls feature interactions between frames. It employs a boundary masking component that constructs a boundary adjacency matrix based on the boundary prediction. This matrix is used to weaken the message passing between frames belonging to different intervals, thereby improving frame-level authenticity decisions.
\end{itemize}

\begin{figure}[htbp]
    \centering
    \includegraphics[scale=0.37]{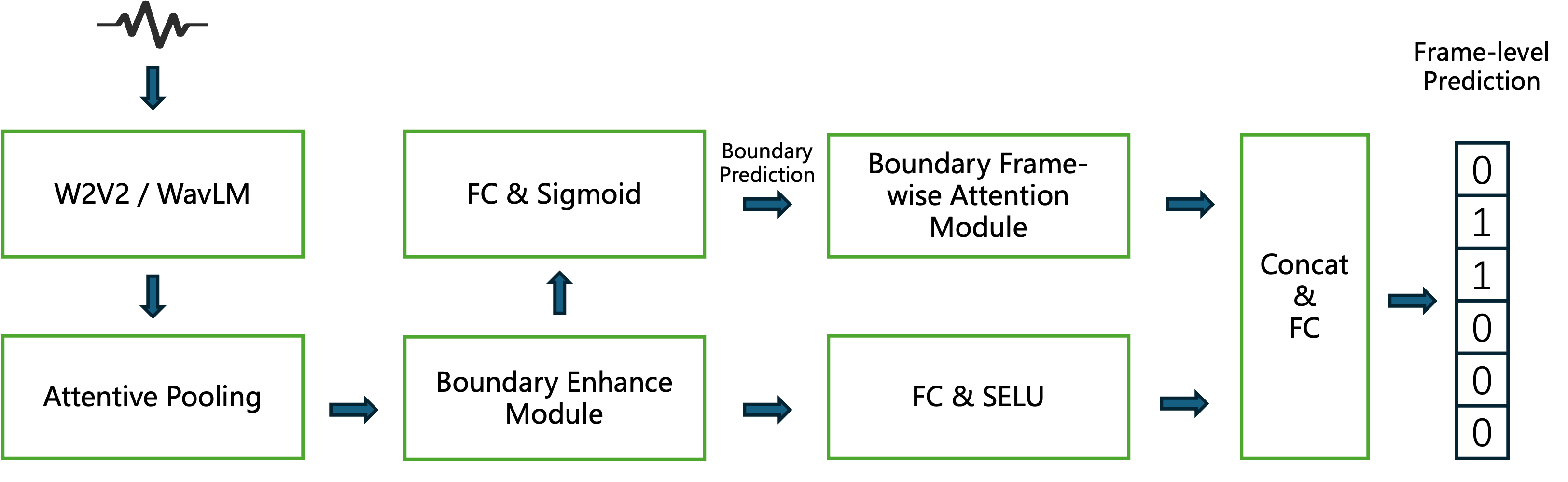}
    \caption{The overall architecture of BAM.}
    \label{fig:bam}
\end{figure}

As illustrated in Figure~\ref{fig:bam}, the raw audio signal is first passed through a front-end feature extractor to obtain the initial acoustic features. These features are then aggregated using attention-based pooling to produce a pooled representation. The pooled features are subsequently fed into BEM and BFAM.

Within the BEM, the features are processed to generate enhanced representations. These enhanced features are used in two ways: one branch passes them through a fully connected (FC) layer followed by a sigmoid activation to predict boundary information; the other branch feeds them into a separate FC layer with a SELU~\cite{klambauer2017self} activation to generate features to be later concatenated with the output of BFAM.

Meanwhile, the predicted boundary information serves as auxiliary input to BFAM, alongside the original pooled features. BFAM processes this combined input to produce interval-aware interaction features. Finally, these features are concatenated with the BEM-generated representations and passed through a fully connected layer to produce the final frame-level prediction results.

During training, we resample all audio data from the official datasets from their original 44.1 kHz to 16 kHz to match the input requirements of our model. For temporal resolution, we adopt 40 ms (instead of the original 160 ms) to align with the minimal forgery granularity in the dataset. Additionally, we fix training samples at 1-second segments, which corresponds to 25 frames per second (FPS), ensuring compatibility with visual-level synchronization when needed. Samples exceeding this length are randomly trimmed, while shorter samples are zero-padded. The data is padded to correspond with the 20 ms shift of WavLM. The model is trained for 50 epochs using the Adam optimizer with an initial learning rate of 1e-5, which is halved every 10 epochs. The final model is selected based on the best validation loss observed during training.

For joint optimization, we use a dual-objective loss function:

\begin{itemize}
    \item Frame-level Authenticity Loss ($L_s$): Standard cross-entropy loss is used for frame-level spoofing detection.
    
    \item Boundary Loss ($L_b$): Binary cross-entropy loss is applied, where ground-truth boundary labels are derived from segment-level authenticity labels. Specifically, only boundary frames are labeled as 1, while all other frames are labeled as 0.
\end{itemize}

The overall loss function is expressed as:

\[
L = L_s(\hat{y}, Y) + \lambda L_b(\hat{b}, B)
\]

Where $\hat{y}$ represents the authenticity prediction result of the model, $Y$ is the frame-level authenticity label, $\hat{b}$ is the boundary prediction probability, $B$ is the ground-truth boundary label, and $\lambda$ is set to 0.5.

\textbf{Inference Strategy.} The model processes variable-length test samples directly, with the boundary predictions determining the final forgery localization results. All other experimental settings remain consistent with the training strategy.

\subsection{Image}
For the video frame branch, we adopt a frame-level detection approach that emphasizes the discriminative capacity of individual frames. This design also serves as a foundation for enabling the model to localize forged segments within the video.

\subsubsection{Training Strategy} 
\textbf{Training Parameter}.
We adopt Xception~\cite{chollet2017xception} as the backbone for both detection and localization. The models are optimized using the Adam optimizer with the following hyperparameters: weight decay = $1\mathrm{e}{-5}$, learning rate = $1\mathrm{e}{-4}$, $\beta_1$ = $0.9$, $\beta_2$ = $0.999$, and $\epsilon$ = $1\mathrm{e}{-8}$. As noted in Sec. ~\ref{sec:data_analysis}, the fake\_audio\_fake\_video subset contains a combination of real and fake frame segments. Given time constraints, training is conducted exclusively on the fake\_audio\_fake\_video subset of the dataset.

\textbf{Data Augmentation}. To improve the robustness of the detection model, we apply data augmentation to the input images with a fixed probability. The specific augmentation strategies are as follows:
\begin{itemize}
    \item Blur: The Gaussian kernel size is randomly chosen from the range [3, 11].
    \item Masking: Random masking of semantically important regions (e.g., facial features such as the eyes, nose, and mouth)
    \item Noise: The noise variance is randomly sampled from the range [10., 60.].
    \item Brightness: The brightness factor ranges from -0.3 to 0.3.
    \item Compression: The compression quality factor (Q) is randomly selected within the range [30, 100].
    \item Resize: A resizing strategy is applied that sequentially performs upsampling by a factor of 1.2, followed by downsampling by 0.8, and then upsampling again by 1.1.
\end{itemize}

\subsubsection{Inference Strategy}
\textbf{Detection}. For detection, the primary challenge lies in determining the authenticity of an entire video based on predictions made at the frame level. The most straightforward approach is to assess video authenticity based on either the proportion of frames predicted as fake or the aggregated confidence scores within a sliding window. However, these simple heuristics are prone to two common pitfalls: excessive sensitivity to isolated fake-frame predictions or an unrealistic bias toward real predictions caused by over-smoothing.

To address this, we adopt a strategy based on the false positive rate (FPR), wherein the model's FPR is first estimated, and video authenticity is then determined by analyzing the relationship between the proportion of frames predicted as fake and the expected FPR.


First, if the number of consecutively forged frames exceeds the maximum number of false positives tolerable by the model—derived from a predefined false positive rate (FPR)—the video is classified as fake (Stage \uppercase\expandafter{\romannumeral1}  $c_1$). However, this criterion primarily targets long forged intervals. As discussed in Sec.~\ref{sec:data_analysis}, many forged segments in fake videos are relatively short (highlighting the risk of prediction smoothing in sliding window–based methods), relying solely on long forged segments may lead to missed detections of more scattered manipulations.

Therefore, attention should also be paid to the relatively sparse and short-duration segments in the prediction results. To address this, we additionally compute the total duration of all forged segments, excluding the longest. If the proportion of this cumulative duration exceeds a threshold lower than the predefined FPR, relative to the total video length, the video is also classified as fake (Stage \uppercase\expandafter{\romannumeral2} $c_2$). Only when neither condition is satisfied is the video considered real. For detailed algorithmic procedures, please refer to Algorithm.~\ref{alg:algorithm_detection}.

\begin{algorithm}[tb]
    \caption{algorithm for detection}
    \label{alg:algorithm_detection}
    \textbf{Input}: video frames $I=\{f_1,f_2,...,f_n\}$\\
    \textbf{Parameter}: model $m(\cdot)$, Stage \uppercase\expandafter{\romannumeral1} constraint $c_1$, Stage \uppercase\expandafter{\romannumeral2} constraint $c_2$, fake frame intervals $F=\{[i,j],[...],[...]\},i<j<n$, confidence score per frame $pred=\{s_1,s_2,...s_n\}$\\
    \textbf{Output}: confidence score $s$
    \begin{algorithmic}[1] 
        \STATE $pred \xleftarrow{} m(I)$
        \STATE $F \xleftarrow{\texttt{IDENTIFY\_FAKE\_INTERVAL}(\cdot)} pred$
        \STATE $F \xleftarrow{} \texttt{REVERSE\_SORT\_BY\_LEN}(F)$
        \IF {$\texttt{LEN}(F[0]) \ge \texttt{LEN}(I) \cdot c_1$}
        \STATE $s \xleftarrow[]{} \texttt{MEAN}(F[0])$
        \ELSIF{$\texttt{SUM}(\texttt{LEN}(F[1:])) \ge \texttt{LEN}(I) \cdot c_2$}
        \STATE $s \xleftarrow[]{} \texttt{MEAN}(F[1:])$
        \ELSE
        \STATE $s \xleftarrow[]{} \texttt{MEAN}(pred)$
        \ENDIF
        \RETURN $s$
    \end{algorithmic}
\end{algorithm}

\textbf{Localization}. For localization, the frame-level design guarantees that predictions for individual frames can be directly aggregated into localization intervals. Consequently, no additional processing is required before fusing these results with the audio modality localization.

\subsection{Fusion}
\subsubsection{Audio-Visual Multimodal Fusion for Detection}
As noted in Sec.~\ref{sec:data_analysis}, the dataset contains cases of unimodal forgeries. Consequently, relying on a single modality for detection is insufficient to fully capture such manipulations, necessitating the fusion of both modalities. 

Due to time constraints, we train the unimodal detection module independently. During inference, each modality-specific module produces predictions on its respective input, and the final decision is derived through fusion. If the predictions from both modules are consistent, we average their confidence scores. If they disagree, to compensate for the limitations of unimodal detection, we bias the decision toward predicting the video as fake and adopt the confidence score from the modality that issued the fake prediction.

\subsubsection{Audio-Visual Multimodal Fusion for Localization}
Considering the limitations of using a single modality for localization, we introduce a temporal interval-based fusion method that effectively integrates complementary information from both the audio and visual modalities along the video timeline.

First, we extract all predicted boundaries from both modalities and partition the entire video into a series of non-overlapping continuous time intervals by removing duplicates and sorting the boundaries. Each time interval corresponds to a segment on the time axis, formally defined as:
\[
T = \{ t_1, t_2, \dots, t_K \}, \quad \text{where } t_1 < t_2 < \cdots < t_K
\]
\[
I_k = (t_k, t_{k+1}), \quad k = 1,2,\dots,K-1
\]
Here, \( t_k \) represents the \( k \)-th timestamp, and \( I_k \) denotes the \( k \)-th time interval.

For each interval \( I_k \), we check whether the audio and visual modality predictions contain segments that overlap with it. If an overlap is found, it indicates that the corresponding modality has detected a forged segment within that interval, and its confidence score is assigned accordingly. If no overlap exists, it means the modality does not provide any valid evidence for that interval, and the confidence score is set to zero.

Then, we apply a max-confidence fusion strategy to combine the confidence scores from the two modalities:
\[
C_{\text{fusion}}(I_k) = \max \left( C_{\text{audio}}(I_k), C_{\text{visual}}(I_k) \right)
\]
where \( C_{\text{audio}} \) and \( C_{\text{visual}} \) denote the confidence scores for the audio and visual modalities within interval \( I_k \), respectively.

In addition, to reduce fragmented detection results and enhance temporal coherence, we further merge consecutive intervals with similar confidence scores. This leads to smoother and more consistent final predictions.

\section{Evaluation}
\subsection{Audio-Visual Multimodal Fusion for Detection}
We validate the effectiveness of our method using evaluation metrics derived from the submitted results. We first obtain prediction results using only the audio detection module, referred to as $w/o$ fusion. Subsequently, we fuse these results with those from the visual detection module, denoted as $w$ fusion. Although the audio-only detection module achieves satisfactory performance, it exhibits certain limitations that impact localization accuracy. By fusing the visual detection results, we observe a 7\% increase in AUC, demonstrating that audio-visual multimodal fusion substantially enhances video authenticity verification. Detailed results are presented in Table.~\ref{tab:fusion-detection}.
\begin{table}[th]
\centering
\begin{tabular}{cccc}
\toprule
\textbf{Fusion Strategy} & \textbf{AUC$\uparrow$} & \textbf{AR$\uparrow$} & \textbf{AP$\uparrow$}\\
\midrule
$w/o$ fusion   & 0.80        & 0.09 & 0.448 \\
$w$ fusion   & 0.872        & 0.23 & 0.551 \\

\bottomrule
\end{tabular}
\caption{Impact of fusion strategy on authenticity judgment. “$w/o$ fusion” denotes results obtained using only the audio modality, while “$w$ fusion” denotes results after fusing both audio and visual modalities. The localization strategy is kept consistent across both cases.}
\label{tab:fusion-detection}
\end{table}

\subsection{Audio-Visual Multimodal Fusion for Localization}
As the precise evaluation scripts for AP and AR calculations were not made available for this track, the fusion performance analysis was necessarily derived from the final benchmark results published by the challenge organizers. Table.  ~\ref{tab:fusion-results}
presents the experimental outcomes of interval-wise score fusion across different modality combinations.

We observe that fusing audio and visual modality confidence scores within defined temporal intervals yields better performance than using either modality alone. Specifically, the combination of pre-trained audio localization and visual localization models improves both Average Precision (AP) and Average Recall (AR) metrics over the single-modality baselines. Furthermore, retraining the audio localization model with additional supervision results in a further performance boost, demonstrating the value of modality-specific fine-tuning.

In particular, under identical detection results, the fused system using a retrained audio model and the visual model achieves the best results, reaching an AP of 0.55 and an AR of 0.23, which are significantly higher than those achieved by any standalone model. This highlights the effectiveness of the proposed interval-wise fusion strategy and its contribution to producing more stable and accurate localization results.
\begin{table}[ht]
\centering
\begin{tabular}{@{}lcc@{}}
\toprule
\textbf{Fusion Strategy} & \textbf{AR} & \textbf{AP} \\
\midrule
Audio Det. Only                        & 0.05 & 0.36 \\
Audio Loc. (Pretrained)               & 0.09 & 0.44 \\
Image Loc. Only                       & 0.13 & 0.39 \\
Audio Loc. (Pretrained) + Image Det.       & 0.13 & 0.50 \\
Audio Loc. (Retrained) + Image Det.        & \textbf{0.23} & \textbf{0.55} \\
\bottomrule
\end{tabular}
\caption{
Localization performance comparison of different fusion strategies based on fixed detection results. 
\textbf{Audio Det.} denotes the detection results produced by the audio module. 
\textbf{Audio Loc. (Pretrained)} represents the localization results produced by a temporal audio localization model pre-trained on the PartialSpoof dataset by its original authors. 
\textbf{Audio Loc. (Retrained)} represents the localization results produced by the same model architecture retrained on the official training set of this challenge. 
\textbf{Image Det.} denotes the detection results produced by the image module. 
\textbf{Image Loc.} represents the localization results produced by the image module.
}

\label{tab:fusion-results}
\end{table}

\section{Conclusions}
In this work, we present a system developed for Track 2 of the Deepfake Detection and Localization Challenge. Our approach leverages different models tailored for different modalities to address the task effectively. By employing an interval-wise fusion strategy, we are able to fully exploit the complementary strengths of audio and visual modalities for both detection and localization.

While our current system adopts a modular and loosely coupled design for flexibility and interpretability, each component is trained and executed independently. In future work, we plan to explore more integrated architectures that jointly optimize across audio and visual streams. Such unified frameworks could reduce implementation complexity while enabling deeper feature-level interactions between modalities. We also aim to enhance the system's ability to detect cross-modal inconsistencies by learning shared temporal and semantic representations. We believe these directions hold great promise for building more efficient, scalable, and generalizable multimodal deepfake detection systems.


\bibliographystyle{named}
\bibliography{ijcai25}

\end{document}